\documentclass[aps,preprint,superscriptaddress]{revtex4-2}

\usepackage{mathtools}
\usepackage{amsmath,bm}
\usepackage{amssymb}
\usepackage{subfigure}
\usepackage{setspace}
\usepackage{dcolumn}
\usepackage{color}
\usepackage{graphicx}
\usepackage{epstopdf}

\usepackage[hidelinks]{hyperref}

\begin{document}

\title{Small-World Disordered Lattices: Spectral Gaps and Diffusive Transport}

\author{Matheus I. N. Rosa}
\affiliation{Department of Mechanical Engineering, University of Colorado Boulder, Boulder CO 80309}
\email{matheus.rosa@colorado.edu, massimo.ruzzene@colorado.edu}
\author{Massimo Ruzzene}
\affiliation{Department of Mechanical Engineering, University of Colorado Boulder, Boulder CO 80309}

\date{\today}

\begin{abstract}
We investigate the dynamic behavior of lattices with disorder introduced through non-local network connections. Inspired by the Watts-Strogatz small-world model, we employ a single parameter to determine the probability of local connections being re-wired, and to induce transitions between regular and disordered lattices. These connections are added as non-local springs to underlying periodic one-dimensional (1D) and two-dimensional (2D) square, triangular and hexagonal lattices. Eigenmode computations illustrate the emergence of spectral gaps in various representative lattices for increasing degrees of disorder. These gaps manifest themselves as frequency ranges where the modal density goes to zero, or that are populated only by localized modes. In both cases, we observe low transmission levels of vibrations across the lattice. Overall, we find that these gaps are more pronounced for lattice topologies with lower connectivity, such as the 1D lattice or the 2D hexagonal lattice. We then illustrate that the disordered lattices undergo transitions from ballistic to super-diffusive or diffusive transport for increasing levels of disorder. These properties, illustrated through numerical simulations, unveil the potential for disorder in the form of non-local connections to enable additional functionalities for metamaterials. These include the occurrence of disorder-induced spectral gaps, which is relevant to frequency filtering devices, as well as the possibility to induce diffusive-type transport which does not occur in regular periodic materials, and that may find applications in dynamic stress mitigation.
\end{abstract}


\maketitle

\section{Introduction}\label{Introduction}
 
Complex networks describe a wide variety of systems in nature and society. In particular, the small-world network model proposed by Watts and Strogatz~\cite{watts1998collective} allows the exploration of the space between regular and random networks. In this model, the vertices of a regular network are re-wired with a probability $p$ to another randomly selected node. This leads to networks that simultaneously exhibit a high degree of clustering, which is characteristic of regular networks, and short vertex-to-vertex distances, which characterizes random networks. Apart from the general characterization of its properties,~\cite{moore2000exact,barrat2000properties,newman2000models,barthelemy2011small,newman2011scaling} the small-world model has been applied to a variety of scenarios such as in the dynamics of epidemic spreading,~\cite{moore2000epidemics,newman2002percolation} and for modeling of brain,~\cite{bassett2006small,liao2017small} social,~\cite{braha2007statistical} and transportation networks.~\cite{latora2002boston,guida2007topology} The physics of condensed matter systems based on small-world networks has also been explored in the form of Ising models,~\cite{herrero2002ising}, to explore the onset of localization~\cite{monasson1999diffusion,zhu2000localization} along with transport in quantum lattices~\cite{kim2003quantum,mulken2007quantum,ccalicskan2007transport}.

In this paper, we investigate the dynamics of simple elastic lattices where non-local connections are inspired by the small-world network model. In the context of phononics and elastic metamaterials,~\cite{hussein2014dynamics} the role of disorder has attracted considerable interest. For example, rainbow-based materials have been investigated for band gap widening and energy trapping~\cite{zhu2013acoustic,cardella2016manipulating,tian2017rainbow,celli2019bandgap,beli2019wave,de2020experimental,thomes2021bandgap}, for wave localization~\cite{flores2013anderson}, to study the occurrence of topological phase transitions~\cite{shi2021disorder} and for signal signal processing applications~\cite{zangeneh2020disorder}. Also, the role of disorder has been explored in the context of phonon and thermal transport~\cite{wagner2016two,hu2019disorder}, and as part of the exploration of quasi-periodic lattices.~\cite{apigo2018topological,rosa2019edge,apigo2019observation,ni2019observation,pal2019topological,
xia2020topological,gupta2020dynamics,rosa2021exploring,rosa2021topological}  So far, most of prior studies have considered disorder introduced to local parameters or couplings, and the introduction of disorder through non-local connections defined by a network model has not yet been explored. To this end, we consider regular mono-atomic spring-mass lattices in the form of 1D lattices and of 2D hexagonal, square and triangular topologies, where  connections are added based on the small-world model characterized by a chosen level of disorder. We first investigate the emergence of spectral gaps as a function of disorder through eigenmode computations performed on large lattices, and for multiple disorder realizations. Robust gaps appear in the form of frequency regions not populated by any modes, or populated only by localized modes, both resulting in low transmission levels across the lattice. These gaps appear to be more pronounced for topologies of lower connectivity, i.e. 1D and 2D hexagonal lattices which exhibit larger gaps than square and triangular lattices. Also, the analysis of transient wave behavior allows the characterization of the transport properties and the identification of transitions from ballistic to super-diffusive and diffusive transport. These are similar to the transitions experienced by quantum and photonic lattices in the presence of on-site disorder~\cite{dunlap1990absence,naether2013experimental}. The investigations presented herein identify a new route for introducing disorder in metamaterials in the form of non-local connections, which holds potential for the generation of disorder-resilient spectral gaps and for applications related to impact mitigation that leverage diffusive transport, as opposed to ballistic spreading observed in periodic materials.

This paper is organized as follows: following this introduction, section~\ref{MainSec} describes the modeling of the small-world phononic lattices and the associated equations of motion. Next, section~\ref{SpecSec} describes the spectral properties of the lattices and the emergence of band gaps through the non-local disordered links. Section~\ref{TransportSec} then describes the transport properties of the network lattices and characterizes the transition from ballistic to diffusive transport. Finally, section~\ref{ConcSec} summarizes the main findings of this work and outlines future research directions. 

\section{Small-world phononic lattices: Description and Equations}\label{MainSec}
\begin{figure}[b!]
\includegraphics[width=1\textwidth]{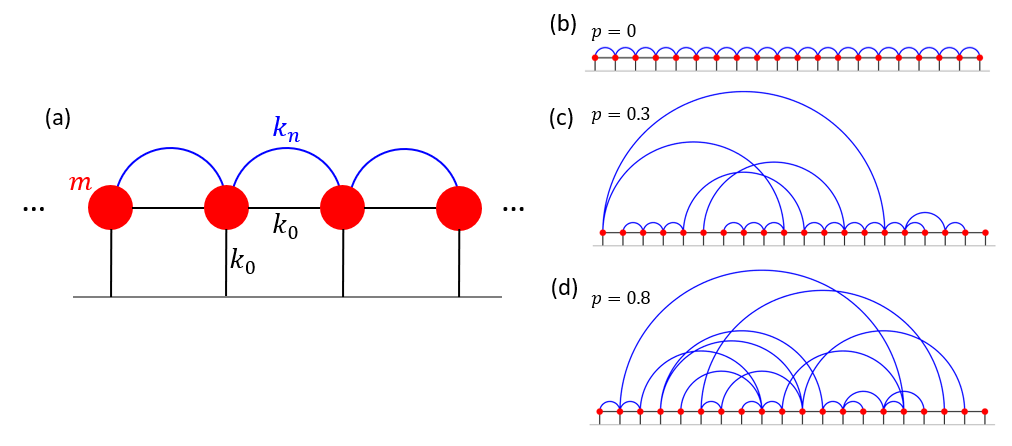}
\caption{Schematics of 1D small-world network lattices. Each mass (red) is connected to the nearest neighbors and to the ground by a spring of constant $k_0$ (black lines) (a). Additional network connections are represented by blue links, which initially also connect nearest neighbors and are re-wired with probability $p$ to another random node. Examples with $N=20$ masses and $p=0,0.3,0.8$ are illustrated in (b,c,d). }
\centering
\label{Fig1}
\end{figure}

\begin{figure}[b!]
\includegraphics[width=1\textwidth]{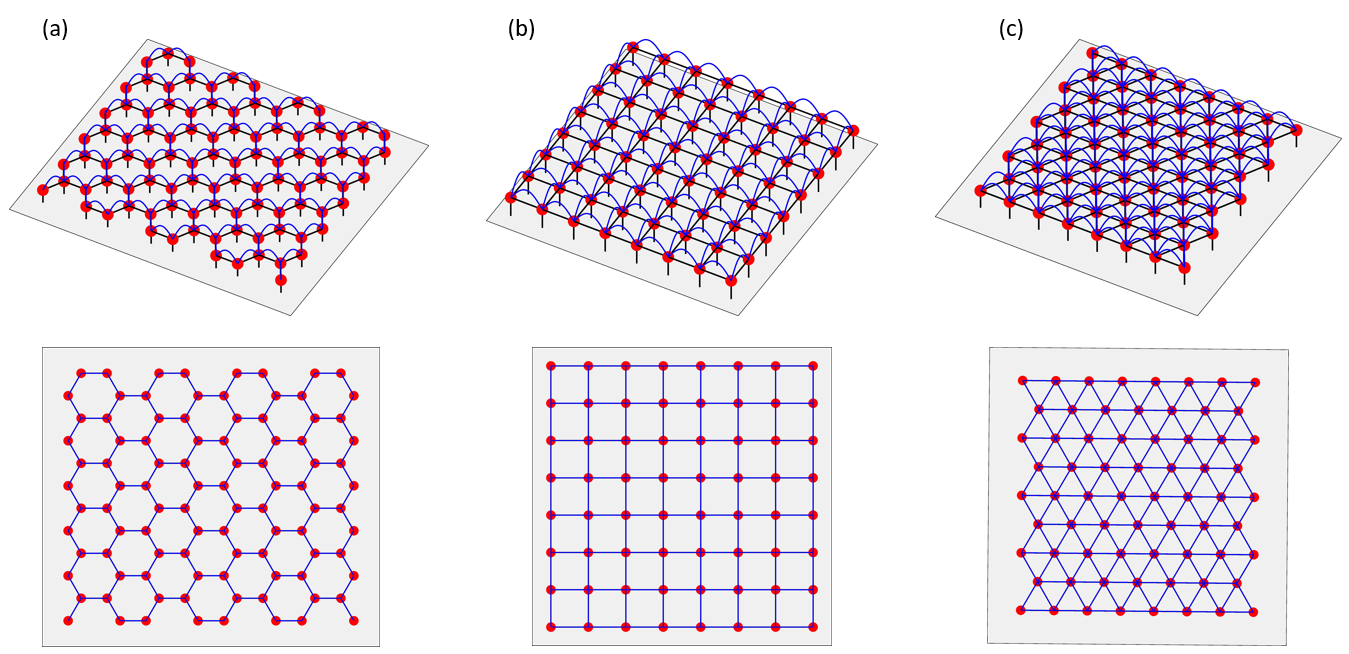}\vspace{3mm}
\includegraphics[width=1\textwidth]{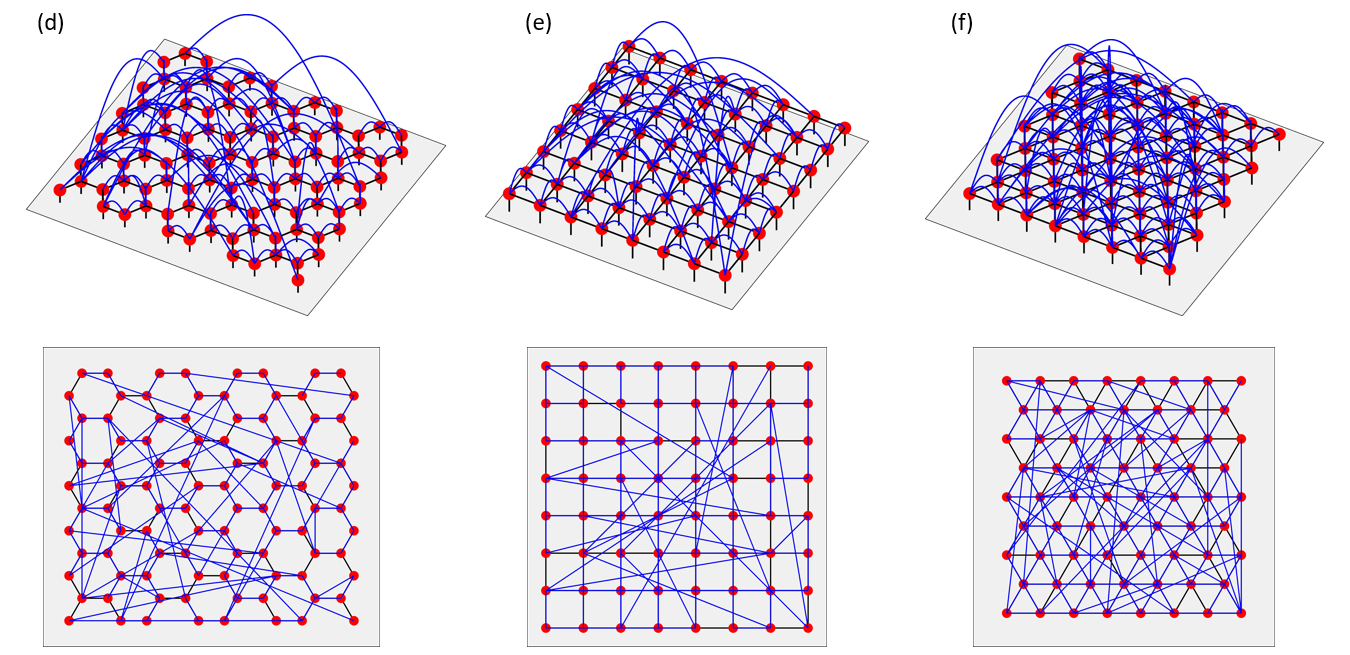}
\centering
\caption{Two-dimensional hexagonal, square and triangular small-world network lattices with $p=0$ (a,b,c) and $p=0.2$ (d,e,f). The top and bottom rows respectively display the perspective and top views, with network links represented by blue lines.}
\label{Fig2}
\end{figure}

Figure~\ref{Fig1} illustrates the considered 1D lattices, which consist of equal masses $m$ (in red) connected to nearest neighbors and to the ground by springs of stiffness $k_0$. These connections form periodic lattices which are here considered as baselines. Additional connections through springs $k_n$, represented as blue lines, connect nodes according to the small-network model based on a probability $p\in [0, \,\, 1]$. The case of $p=0$ defines additional nearest neighbor connections (Figs.~\ref{Fig1}(a,b)), while $p\in (0, \,\, 1]$ in general defines the probability of rewiring to randomly chosen nodes (Figs.~\ref{Fig1}(c,d)). The same approach is employed for the 2D lattices in Fig.~\ref{Fig2}, which shows examples of hexagonal, square and triangular lattices with $p=0$ (a,b,c), and with $p=0.2$ (d,e,f). 
The presence of an underlying lattice guarantees that all the masses are connected, which is not enforced by the small-world network (blue) alone for arbitrary values of $p$.~\cite{watts1998collective} Also, we elect that the spring constants are inversely proportional to the distance, i.e $k_{n,s}=\alpha k_0/d_{n,s}$, where $d_{n,s}$ denotes the distance separating masses indexed by $n$ and $s$, while $\alpha$ defines indicates the strength of these connections compared to the underlying lattice. This choice is motivated by the desire to retain the physical behavior of couplings in a mechanical lattice, whereby stiffness terms are typically inversely proportional to the length of the connection.

The equation of motion for a mass of index $n$ is expressed as:
\begin{equation}
m\ddot{u_n}+k_0u_n+\sum_{r}k_0(u_n-u_r)+ \alpha\sum_{s}\frac{k_0}{d_{n,s}}(u_n-u_s)=f_n,
\end{equation}
where $r$ runs over the nearest neighbors, and $s$ runs over the network links that connect to that mass, while $u_n$ and $f_n$ represent the displacement and the force applied to the $n-th$ mass. For a finite lattice of $N$ masses, the equations of motion can be assembled in matrix form:
\begin{equation}
\mathbf{M}\ddot{\mathbf{u}}+\mathbf{K}\mathbf{u}=\mathbf{f},
\end{equation}
where $\mathbf{u}=[u_1, u_2, ..., u_N]^T$, $\mathbf{f}=[f_1, f_2, ..., f_N]^T$, and $\mathbf{M}$,$\mathbf{K}$ respectively denote the mass and stiffness matrices. The numerical results presented in this paper rely on standard procedures such as numerical solution of the eigenvalue problem $\mathbf{Ku}=\omega^2\mathbf{Mu}$ for the natural frequencies and mode shapes of the lattice, and numerical integration of the equations of motion for evaluating the transient behavior under a set of initial conditions, both in the absence of external forcing. Additionally, we evaluate the lattice response for assigned harmonic forcing $\mathbf{f}(t)=\mathbf{f_0}e^{i\omega t}$ by numerically solving the linear system $(-\omega^2\mathbf{M}+\mathbf{K})\mathbf{u}=\mathbf{f_0}$, with $\mathbf{f_0}$ being the vector of forcing amplitudes at frequency $\omega$.

\begin{figure}[b!]
\includegraphics[width=1\textwidth]{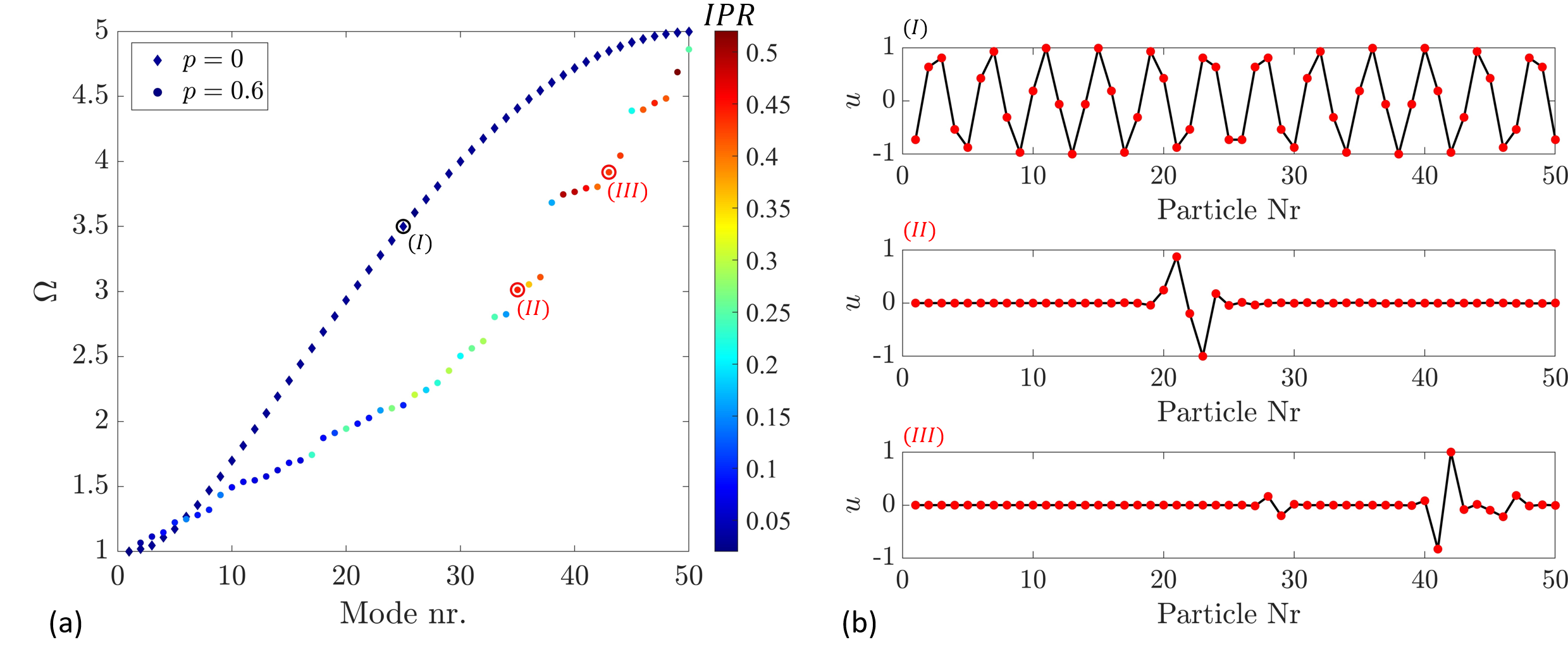}
\caption{Eigenfrequencies (a) and selected mode shapes (b) of finite lattice with $N=50$ masses and free-free boundary conditions for $p=0$ and $p=0.6$. The frequencies in (a) are color-coded according to their IPR, signaling whether the modes are localized or not.}
\centering
\label{Fig3}
\end{figure}

\section{Spectral properties and band gap emergence}\label{SpecSec}
We first investigate the spectral properties of the network lattices and the emergence of band gaps. Due to the absence of periodicity, we rely on eigen-mode computations performed on representative finite lattices. Figure~\ref{Fig3}(a) illustrates an example for a finite lattice with $N=50$ masses and $\alpha=5$, and compares the eigen-frequencies computed for two lattices with $p=0$ and $p=0.6$ under free-free boundary conditions. Throughout this paper, we employ $\Omega=\omega/\omega_0$ as a normalized frequency, where $\omega_0=\sqrt{k_0/m}$. The dots corresponding to each frequency value are color-coded based on the Inverse Participation Ratio (IPR) of the corresponding modes, which is defined as:
\begin{equation}
IPR=\frac{\sum_n |u_n|^4}{\left( \sum_n |u_n|^2 \right)^2},
\end{equation}
where $u_n$ is the $n_{th}$ component of the eigenvector. The IPR varies from 0 to 1 and signals whether a mode is localized or not when its value is high or low, respectively. A few modes are marked in Fig.~\ref{Fig3}(a) and have their mode shapes displayed in Fig.~\ref{Fig3}(b). When $p=0$, the frequencies of the lattice define a continuous band with only non-localized modes, as expected of a periodic monoatomic lattice~\cite{hussein2014dynamics}. However, the lattice with $p=0.6$ supports a series of localized modes, with two examples displayed in Fig.~\ref{Fig3}(b). While localized modes are expected to appear due to the presence of disorder, we also note that a few frequency bands are not populated by any modes and may potentially define band gaps. Naturally, these results correspond to a single realization of the lattice, and are not sufficient to draw conclusions about the behavior expected of any realization due to the randomness of the non-local connections.

\begin{figure}[b!]
\begin{subfigure}[]{
\includegraphics[width=1\textwidth]{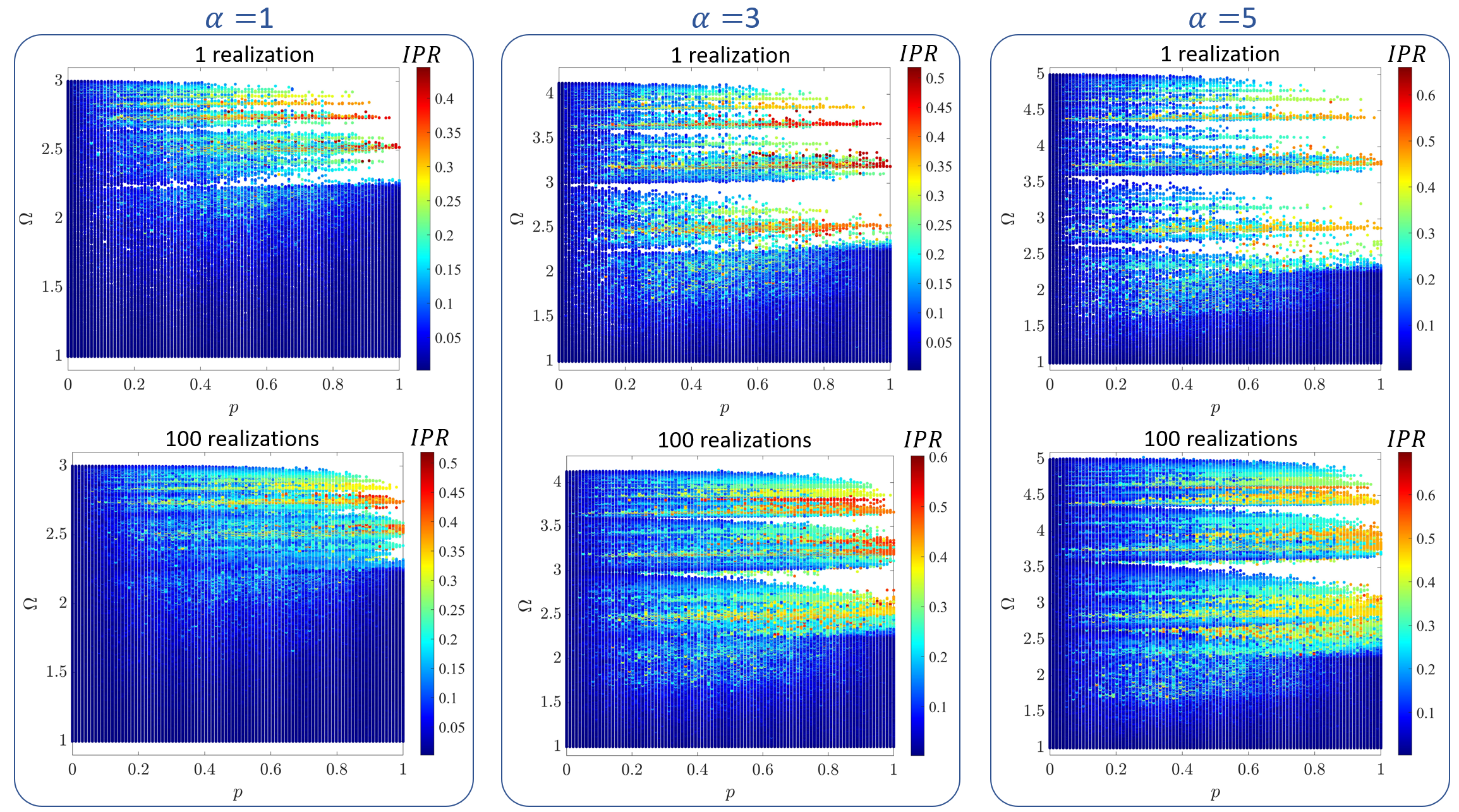}}
\centering
\end{subfigure}
\begin{subfigure}[]{
\includegraphics[width=1\textwidth]{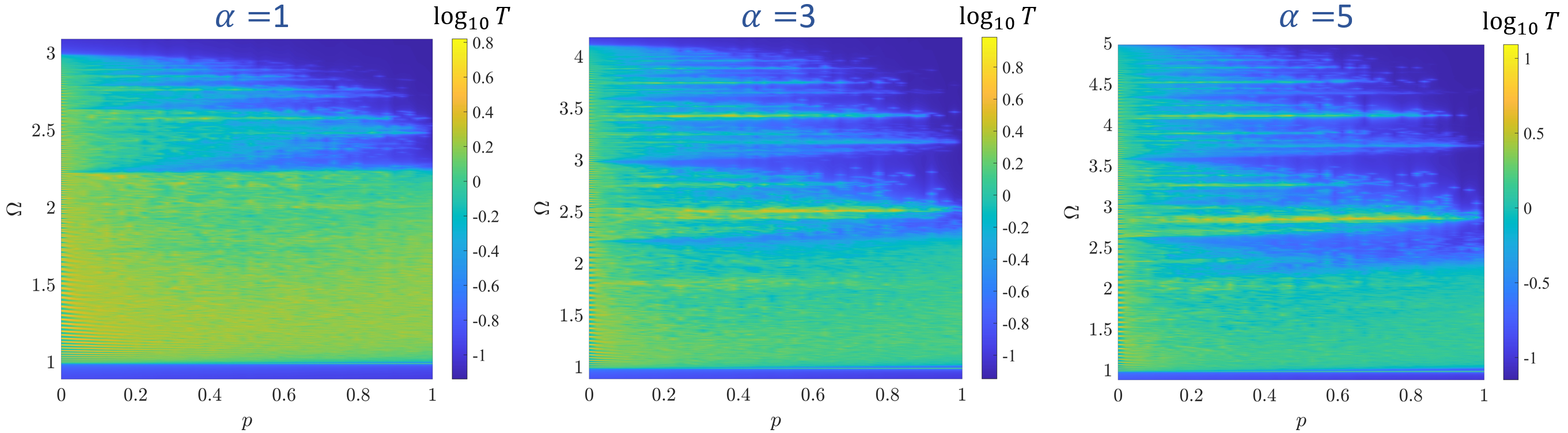}}
\centering
\end{subfigure}
\caption{Spectral properties of 1D small-world phononic lattices. Eigenfrequencies of finite lattice with $N=500$ masses computed as a function of $p$ for multiple $\alpha$ values and $1$ (top) or 100 (bottom) lattice realizations, color coded by the IPR (a). Transmission of finite lattice with $N=200$ masses when excited at the center as a function of $p$, averaged along the lattice and across 100 realizations, for multiple $\alpha$ values (b).}
\centering
\label{Fig4}
\end{figure}

To obtain further insight into the spectral properties of the network lattices, we compute the modes of a large finite lattice comprising $N=500$ masses for $p$ varying from $0$ to $1$, for multiple realizations and for multiple values of $\alpha$, with results summarized in Fig.~\ref{Fig4}(a). Each column corresponds to one $\alpha$ value ranging from $1$ to $5$:  the top row displays the frequencies as a function of $p$ for a single lattice realization, while the bottom row superimposes the frequencies of $100$ different random realizations of each $p$ value. In both cases, the color is associated with the IPR. The results confirm the existence of frequency bands where no modes exist, which emerge for increasing values of $p$, and are more pronounced for larger values of $\alpha$. The presence of these spectral gaps is further confirmed by evaluating the harmonic response of a lattice with $N=200$ masses, excited by a harmonic force at the center mass. The harmonic response is computed as described in section~\ref{MainSec}, and we define the transmission as the $\mathcal{L}_2$ norm of the response along the lattice divided by the response at the input site $n_0$, i.e. $T=||\mathbf{u}/u_{n0}||_2$. The transmission is averaged across 100 realizations for varying $p$ and $\Omega$, and is displayed as log-scale colormaps in Fig.~\ref{Fig4}(b) for the different $\alpha$ values. We note that for $\alpha=1$ a few gaps appear for a single realization, which are mostly filled by localized modes characterized by a high IPR when multiple realizations are considered. For $\alpha=3$ and $\alpha=5$, some gaps are persist without any existing modes even for multiple realizations, signaling a degree of robustness. The transmission results in Fig.~\ref{Fig4}(b) confirm the existence of the spectral gaps that are not occupied by any modes, and also of the gaps that are populated by a high density of localized modes. Indeed, localized modes are not associated with transport along the lattice, and therefore are associated with frequency regions of low vibration transmission along the length of the lattice. These results illustrate that the addition of small-world networks of springs of sufficient strength produce well defined spectral gaps that emerge due to the disordered network connections, and persist across multiple lattice realizations. We also note the presence of thin bands of high transmission at fixed frequencies spanning large $p$ intervals, for example at $\Omega=2.5, 3.4$ with $\alpha=3$, and $\Omega=2.8, 4.2$ with $\alpha=5$, which signal another robust feature emanating from such disordered lattices. 

\begin{figure}[b!]
\includegraphics[width=1\textwidth]{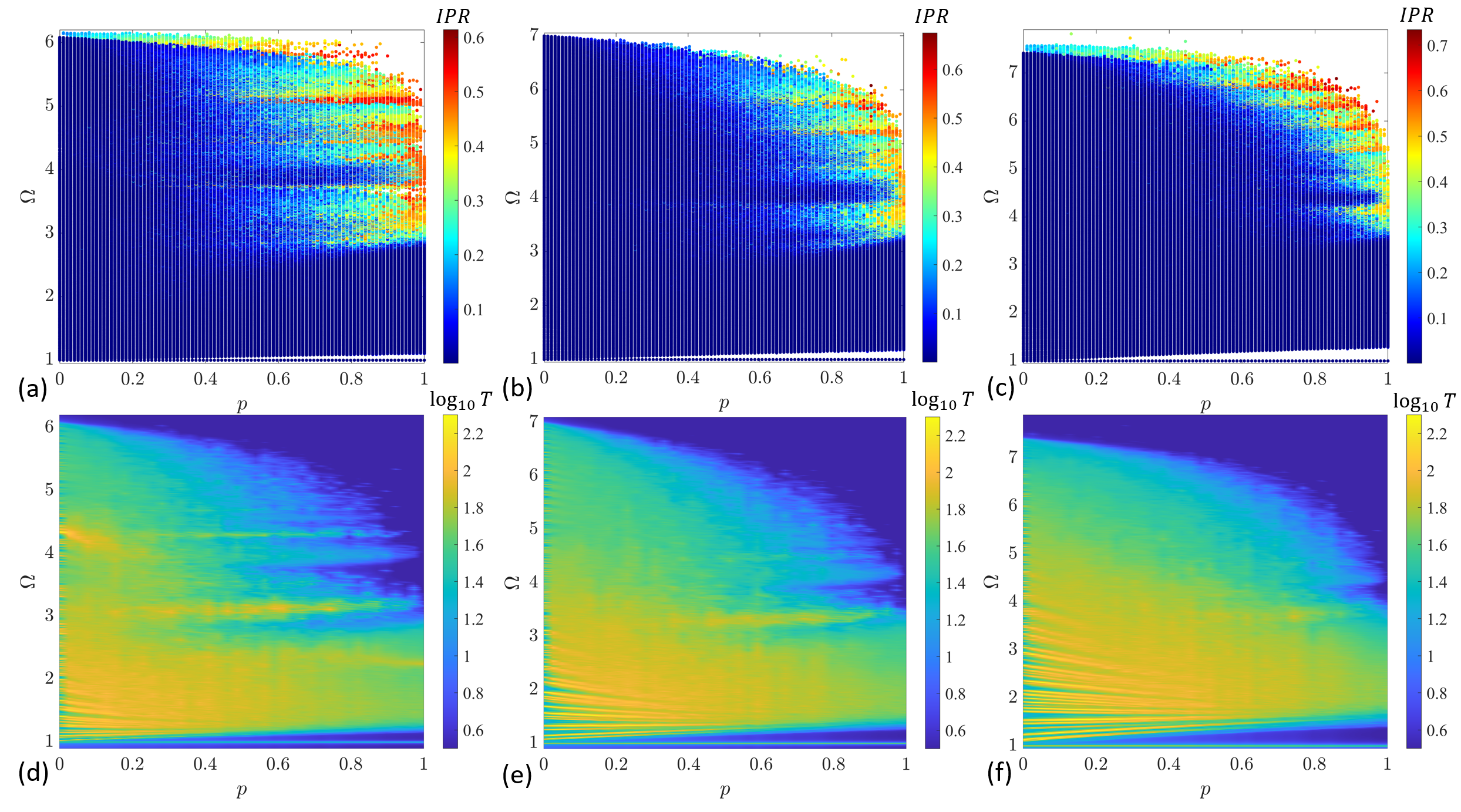}
\caption{Spectral properties of 2D network lattices. Top panels (a,b,c) display the eigenfrequencies of $61 \times 61$ hexagonal, square and triangular lattices, considering 100 realizations, and color-coded according to the IPR. The bottom panels (d,e,f) display the transmission averaged over 100 realizations for $41 \times 41$ hexagonal, square and triangular lattices.}
\centering
\label{Fig5}
\end{figure}

Next, we investigate the spectral properties of 2D lattices. Results for $\alpha=5$ are summarized in Fig.~\ref{Fig5}, which displays the frequencies of $61 \times 61$ lattices as a function of $p$. As in previous figures, the spectral properties are color-coded in terms of the IPR, and are obtained considering 100 different realizations of hexagonal (a), square (b) and triangular (c) lattices. The bottom panels (d,e,f) display the transmission as a function of $p$ and $\Omega$, which are obtained for $41 \times 41$ lattices excited at the center mass and by averaging the response of 100 lattice realizations. While no regions that are not populated by any modes are found, there is evidence of regions populated only by localized modes emerging with increasing $p$. As in the 1D case, these regions filled by localized modes correspond to regions of low transmission, as confirmed by the plots in (d,e,f). In the hexagonal case, a couple of regions within the specturm filled with localized modes are observed to emerge in Fig.~\ref{Fig5}(a), which are confirmed by low transmission regions in (d). The size of the gaps within the spectrum are diminished in the square lattice case (b,e), and become very small in the triangular lattice case (c,f). These results illustrate how the lattice tolology plays an important role in the emergence of bandgaps, and suggest that the progressive increase in connectivity from the 1D lattice (2 connections per node) to the triangular lattice (6 connections per node) causes a decrease in band gap occurance. In particular to the 2D case, we note that low frequency gaps seem to emerge as a function of $p$ as the first mode of the lattice is separated by a gap from the collective of the other modes of the lattice. The width of this gap increases with the lattice connectivity, i.e. it appears wider in triangular lattices than in the hexagonal ones. Although the spectral properties of these lattices are largely influenced by the topology, in the next section we illustrate that the transport properties are qualitatively similar and that transitions to diffusive transport can be observed in all cases.

\section{Transient behavior: from ballistic to diffusive transport}\label{TransportSec}

The transient behavior of the small-world lattices is investigated next. We characterized wave motion in disordered lattices by relying on a approach~\cite{tang2018experimental,naether2013experimental} that considers the dynamic evolution of the lattice motion resulting from an initial perturbation. Such evolution is quantified by computing the Mean Square Displacement (MSD), which is defined as
\begin{equation}
MSD(t)=\left< \sum_n (d_{n,n_0})^2|u_n(t)|^2 \right> \approx t^\gamma,
\end{equation}

Here, $\left< \right>$ denotes the averaging operation across multiple lattice realizations, $n_0$ is the site where the initial perturbation is applied, and $d_{n,n_0}$ is the distance between the generic site $n$ and $n_0$. The MSD is found to scale as $t^\gamma$, where $t$ denotes time while the exponent $\gamma$ quantifies the rate of perturbation spreading. Thus, $\gamma$ is used to classify the type of transport that occurs along the medium. For example, quantum and photonic periodic lattices, which are governed by similar equations of motion, exhibit \emph{ballistic} transport  which is characterized by $\gamma=2$ ~\cite{tang2018experimental,naether2013experimental}. In the presence of disorder, decreasing values of $\gamma$ quantify the slower spread that occurs in comparison to regular periodic materials. For instance, depending on the amount of disorder, lattices may exhibit super-diffusive ($\gamma=1.5$) or diffusive ($\gamma=1$) transport, or absence of transport altogether for $\gamma\approx0$, which corresponds to the onset of Anderson Localization~\cite{anderson1958absence,dunlap1990absence,naether2013experimental}. This approach has been recently applied to other types of aperiodic systems, for example in fractal lattices where $\gamma$ is found to be related to the fractal dimension of the lattice,~\cite{xu2021quantum} and also for the characterization of wave packets spreading in disordered non-linear architected materials~\cite{ngapasare2022wave}.

Here, we adopt the MSD to characterize the transport in the small-world lattices. Starting from the 1D case, we consider a large lattice with $N=1000$ masses and apply a perturbation to the $n_0=500$ site. The perturbation is in the form of initial conditions $u_{n_0}(0)=0, \dot{u}_{n_0}(0)=1$, which are equivalent to an impulse excitation $f(t)=\delta(t)$ applied to the chosen site. This excitation involves the entire spectrum of the lattice and is similar to the excitation applied to photonic lattices at $z=0$, where $z$ is the propagation dimension~\cite{naether2013experimental}. We observe the spreading for a series of 1D lattices of different $\alpha$ values and evaluate its variation in terms of the disorder parameter $p$. The simulation time window is adjusted for each $\alpha$ based on the $p=0$ case in order to avoid the presence of reflections at the boundaries, which would affect the MSD computations. For each $\alpha,p$ combination, the MSD is computed by averaging across $200$ lattice realizations, and the resulting $\gamma$ is extracted by fitting the tail of the corresponding $MSD(t)$ curve in logarithmic scale.

\begin{figure}[b!]
\includegraphics[width=1\textwidth]{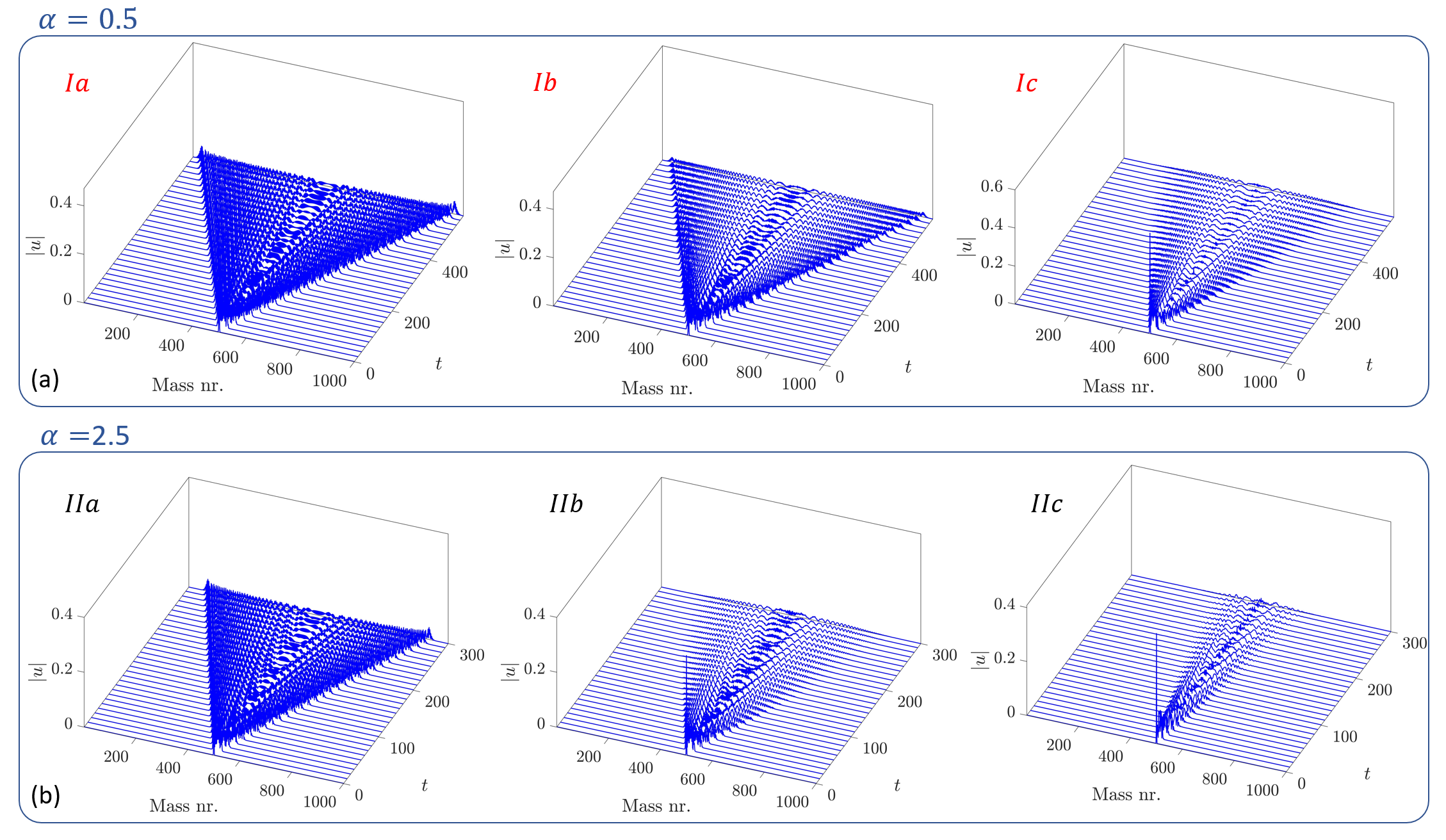}
\includegraphics[width=1\textwidth]{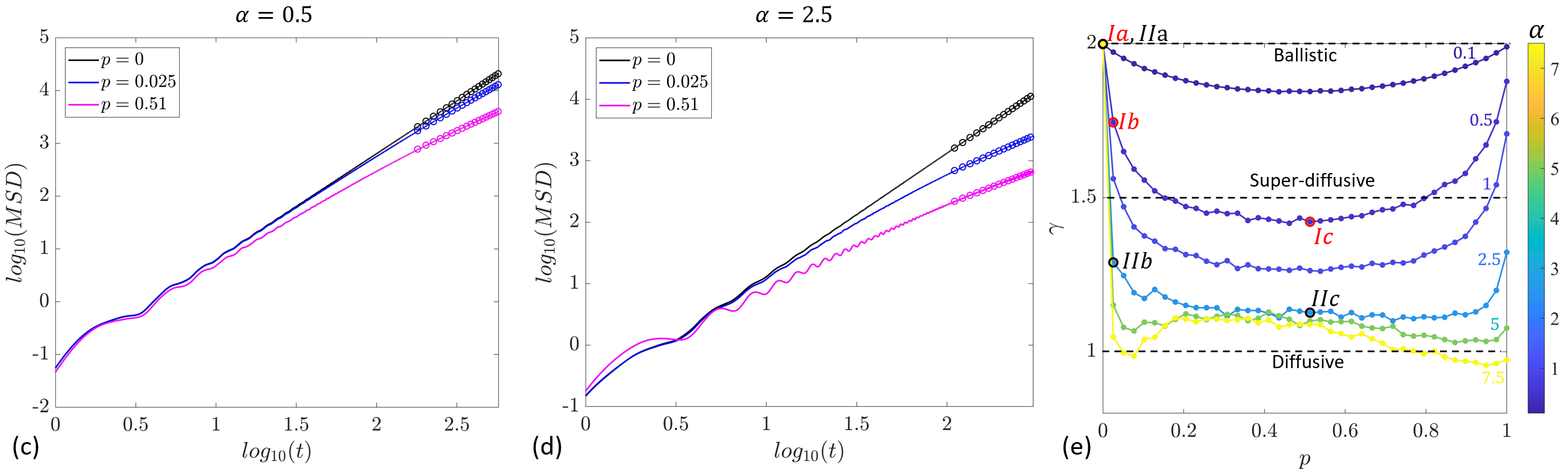}
\centering
\caption{Transport properties of 1D small-world network lattices. Average displacement fields for $\alpha=0.5$ (a) and $\alpha=2.5$ (b), with $p=0$ (left), $p=0.025$ (middle) and $p=0.51$ (right). The fitting of the corresponding MSD curves is displayed in (c,d), with circles superimposed to the curves corresponding to fitted data. Variation of $\gamma$ with $p$ for multiple $\alpha$ values (e), with guidelines for ballistic, super-diffusive and diffusive transport.}
\label{Fig6}
\end{figure}

The results for the 1D lattices are summarized in Fig.~\ref{Fig6}.  Figures~\ref{Fig6}(a,b) display the average displacement field in absolute value for $\alpha=0.5$ and $\alpha=2.5$, and for increasing values of $p$. The results illustrate a decrease in the spreading of the perturbation caused by the increase in disorder from the baseline $p=0$ case (left panels), as $p$ increases to $p=0.025$ (middle panels), and up to $p=0.51$ (right panels). The quantification of the spreading using the $MSD$ is illustrated in Figs.~\ref{Fig6}(c,d) by the plots in log scale. The circles superimposed to the tail of the $MSD$ curves correspond to the numerically fitted relationship $t^\gamma$, allowing for the extraction of $\gamma$ that quantifies the observed decrease in slope associated with the decrease in spreading. The exemplified procedure is repeated for multiple lattices resulting in Fig.~\ref{Fig6}(e), where the extracted $\gamma$ values are directly plotted as a function of $p$ for multiple $\alpha$ values ranging from $0.5$ to $7.5$. The figure provides a characterization of the transport properties of the lattice, illustrating that multiple transport regimes such as ballistic, super-diffusive and diffusive are achieved by different parameters $p,\alpha$. In particular, we confirm that for $p=0$, ballistic spreading characterized by $\gamma=2$ occurs regardless of the $\alpha$ value, as expected from periodic lattices. As $p$ increases, $\gamma$ decreases along different transitions that are intensified for higher values of $\alpha$. The results in Figs.~\ref{Fig6}(a,b) correspond to the points marked in Fig.~\ref{Fig6}(e), chosen to illustrate two  transitions types. The one in Fig.~\ref{Fig6}(a) corresponds to a transition from ballistic to super-diffusive transport as $\gamma$ approaches $1.5$ for increasing disorder levels. The second example in Fig.~\ref{Fig6}(b) illustrates the transition from ballistic to diffusive transport as $\gamma$ becomes closer to $1$. 

\begin{figure}[b!]
\includegraphics[width=1\textwidth]{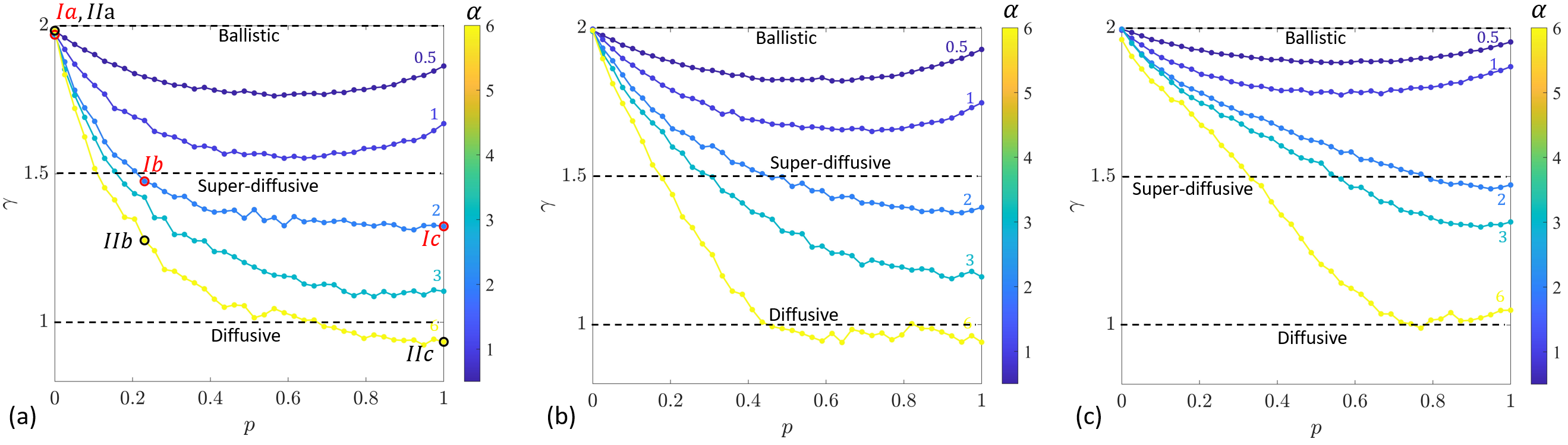}
\includegraphics[width=1\textwidth]{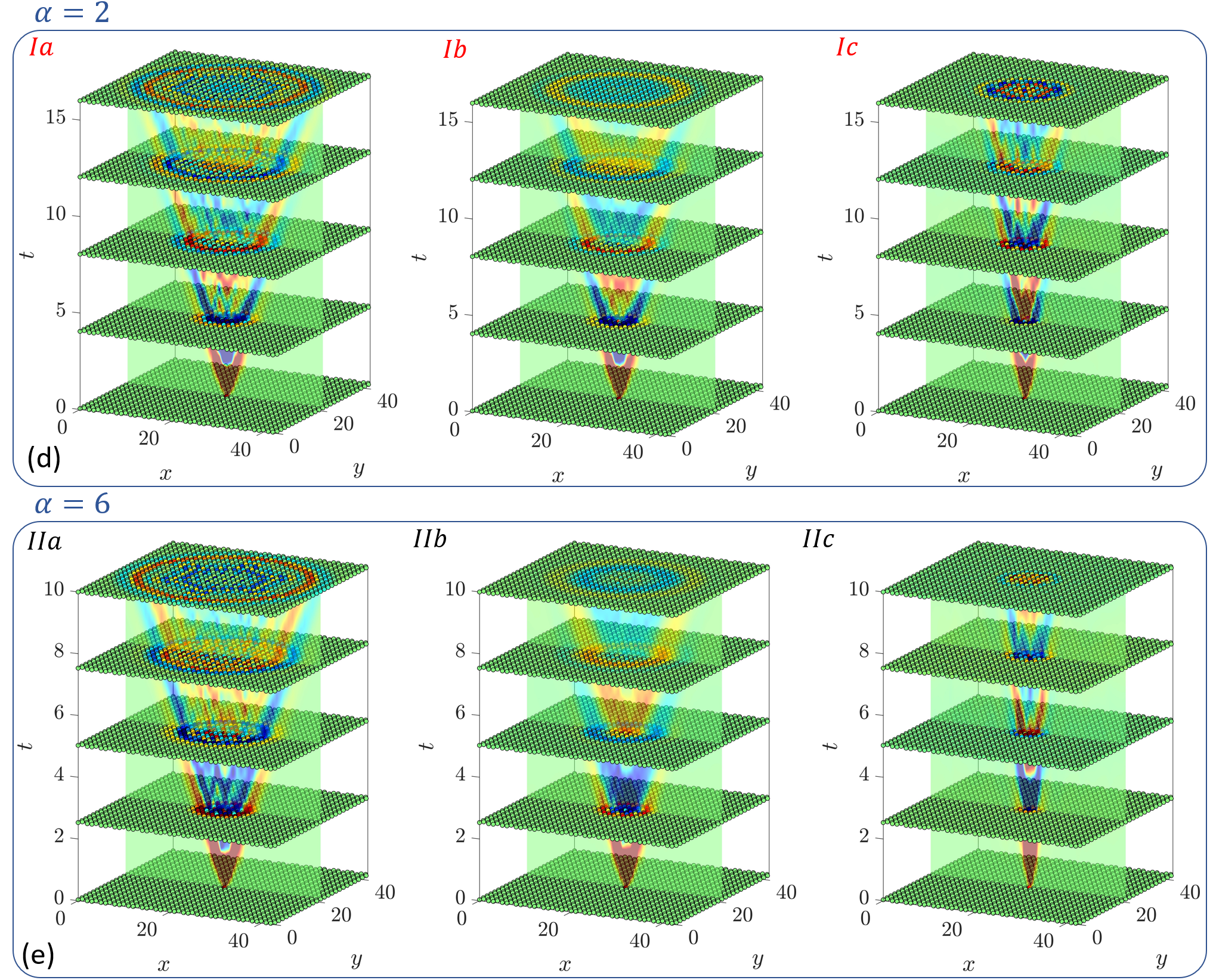}
\centering
\caption{Transport properties of 2D small-world network lattices. Variation of $\gamma$ with $p$ for hexagonal (a), square (b) and triangular (c) lattices, and for multiple $\alpha$ values ranging from $0.5$ to $6$. Examples marked in (a) have their corresponding averaged displacements displayed in (d,e).}
\label{Fig7hexag}
\end{figure}

Next, the transport behavior for the 2D hexagonal, square and triangular lattices is illustrated in Fig.~\ref{Fig7hexag}. The procedure for the 1D case is extended to 2D lattices of sizes $29 \times 52$, $41 \times 41$ and $49 \times 57$ , respectively. The number of cells has been chosen in order to form domains of similar length along the $x$ and $y$ directions, and which are sufficiently large to observe the spreading of the perturbation applied as initial conditions to the center mass. Similar to the 1D case, the results are obtained upon averaging again over $200$ realizations and extracting $\gamma$ from the $MSD$ curves for each $p,\alpha$ combination. Figures~\ref{Fig7hexag}(a,b,c) display the resulting variation of $\gamma$ with respect to $p$ for the hexagonal, square and triangular lattices, with $\alpha$ ranging from $0.5$ to $6$. As in the 1D case, we again find that periodic lattices for $p=0$ are characterized by ballistic transport properties associated with $\gamma=2$. This is verified for any value of $\alpha$ for all the lattice topologies considered, and it is expected for periodic lattices in general.~\cite{tang2018experimental} The figures also illustrate how multiple transport regimes are achieved in all lattice topologies by different choices of $p,\alpha$. Two transitions with $\alpha=2$ and $\alpha=6$ are exemplified for the hexagonal lattice, where the points marked in Fig.~\ref{Fig7hexag}(a) have their corresponding averaged displacement fields displayed in panels (d,e). The plots show the absolute value of displacement across the lattice in the $x,y$ plane at 5 subsequent time instants, with time varying along the vertical axis. The displacement for a sectional $x,t$ plane defined for the center $y$ coordinate is also plotted to improve the visualization of the wave spreading as a function of time. Also, due to the amplitude decrease resulting from wave spreading, the color axis in each plot is restricted to a range corresponding to $10\%$ of the maximum displacement value along the entire time history. For $\alpha=2$, a transition to super-diffusive behavior ($\gamma \approx 1.5$) is observed, while $\alpha=6$ produces a transition to diffusive transport ($\gamma \approx 1$). The associated decrease in the spreading can be clearly observed in the displacement plots by observing how the wave front propagates shorter distances in the $p=1$ cases when compared to the ballistic $p=0$ cases. Overall, the transport transitions are very similar for all the considered 2D lattice topologies, with small qualitative differences in the $\gamma$ variations with $p$ and $\alpha$.

These results illustrate how the disorder introduced through the network connections modify the type of transport for all the considered lattice topologies, causing a transition to super-diffusive or diffusive transport when the strength of the network connections ($\alpha$) is sufficiently strong. These transitions are reminiscent of those experienced by quantum and photonic lattices with on-site disorder~\cite{dunlap1990absence,naether2013experimental}. However, we note that for the considered range of $\alpha$ values, $\gamma$ reaches a plateau close to $1$, and Anderson Localization ($\gamma \approx 0$) does not occur. Higher values $\alpha>7.5$ are not considered herein since the network connections become much stronger and overcome the couplings of the underlying lattice. Our preliminary investigations showed that the transport in that case was not well captured by the MSD computations, similarly to findings in quantum lattices with distance-independent coupling and absence of an underlying lattice~\cite{kim2003quantum}. Such high $\alpha$ regime may be further investigated in the future.

\section{Conclusions}\label{ConcSec}
In this paper, we investigate the dynamics of phononic lattices with small-world network connections. Our results illustrate the emergence of spectral gaps due to increasing degrees of disorder, which are persistent across multiple lattice realizations. Lattices of different topologies, such as 1D and 2D hexagonal, square and triangular lattices were shown to feature transitions from ballistic to super-diffusive or diffusive transport. These results motivate a new route for the introduction of disorder in metamaterials through network connections, potentially leading to novel functionalities enabled by disorder such as spectral gaps and diffusive transport, which could be exploited in impact mitigation applications for example. The initial investigations presented here may be expanded in multiple directions in future studies. For example, a variety of other network modeling strategies may be considered, along with different underlying lattice topologies, different statistical modeling of non-local connections instead of random re-wiring, the introduction of non-linearities, as well as the experimental investigations of the transport properties, among others. 

\begin{acknowledgments}
The authors gratefully acknowledge the support from the National Science Foundation (NSF) through the EFRI 1741685 grant and from the Army Research office through grant W911NF-18-1-0036.
\end{acknowledgments}

\bibliographystyle{unsrt}
\bibliography{References}

\end{document}